\documentclass{aastex}          
\usepackage{spr-astr-addons}    

\usepackage{latexsym,comment}
\usepackage{amssymb}
\usepackage{amsfonts}
\usepackage{amsmath,color}

\def\mb#1{\mathbf{#1}}

\def\eqi{\begin{equation}}
\def\eqf{\end{equation}}
\def\eqI{\begin{equation}}
\def\eqF{\end{equation}}
\def\bds#1{\boldsymbol{#1}}
\def\eqIa{\begin{eqnarray}}
\def\eqFa{\end{eqnarray}}

\def\ber{\begin{eqnarray}}
\def\eer{\end{eqnarray}}
\def\beq{\begin{equation}}
\def\eeq{\end{equation}}

\def\rmd{{\rm d}}

\def\dal{\mathop{\rlap{\hbox{$\sqcap$}}\sqcup}\nolimits}   
\newcommand{\ppar}[2]{\frac{\partial #1}{\partial #2}}

\begin{document}
%
\title{Gravitomagnetic Field of Rotating Rings}

\shorttitle{Gravitomagnetic Field of Rotating Rings}
\shortauthors{<Autors et al.>}

\author{Matteo Luca Ruggiero\altaffilmark{1}} 

\altaffiltext{1}{DISAT, Politecnico di Torino, Corso Duca degli Abruzzi 24, Torino, Italy\\
 INFN, Sezione di Torino, Via Pietro Giuria 1, Torino, Italy}

\begin{abstract}
In the framework of the so-called gravitoelectromagnetic formalism, according to which the equations of the gravitational field can be written in analogy with  classical electromagnetism, we study the gravitomagnetic field of a  rotating ring, orbiting around a central body.  We calculate the gravitomagnetic component of the  field, both in the intermediate zone between the ring and the central body, and far away from the ring and central body. We evaluate the impact of the gravitomagnetic field on the motion of test particles and, as an application, {we study the possibility of using these results, together with the Solar System ephemeris, to infer information on the spin of ring-like structures.}
\end{abstract}

\section{Introduction}\label{sec:intro}

In General Relativity (GR) mass currents give rise to gravitomagnetic (GM) fields, in analogy with classical electromagnetism: actually, the field equations of GR, in linear post-newtonian approximation, can be written in form of Maxwell equations for the gravitoelectromagnetic (GEM) fields\citep{2002NCimB.117..743R,2001LNP...562...83M}, \citep{2003gr.qc....11030M}, where the gravito-electric (GE) field is just the Newtonian field. Even though these effects are normally very small and hard to detect, there have been many efforts to measure them. For instance, the famous Lense-Thirring effect \citep{1918PhyZ...19..156L},  that is the precessions of the node and the periapsis of a satellite orniting a central spinning mass, has been analyzed in different contexts: there are the LAGEOS tests around the Earth \citep{ciufolini2004confirmation,ciufolini2010gravitomagnetism}, the MGS tests around Mars \citep{2006CQGra..23.5451I,2010CEJPh...8..509I} and other tests around the Sun and the planets \citep{2012SoPh..281..815I};  see  \citet{ciufolini2007dragging,2011Ap&SS.331..351I,2013AcAau..91..141I,renzetti2013history} for a discussion and a review of the recent results. In February 2012 the LARES mission \citep{2012EPJP..127..133C} has been launched to measure the Lense-Thirring effect of the Earth, and is now  gathering data; a comprehensive discussion on this mission can be found in  \citet{2005NewA...10..616I,2009SSRv..148..363I,renzetti2013first,renzetti2012higher,ciufolini2010lares,ciufolini2015preliminary}. In the recent past, the Gravity Probe B \citep{2011PhRvL.106v1101E} mission  was launched to measure the precession of orbiting gyroscopes   \citep{pugh1959proposal,schiff1960possible}. The GM clock effect, that is the difference in the proper  periods of standard clocks in prograde and retrograde circular orbits around a rotating mass, has been investigated  but not detected yet \citep{mashhoon2001gravitomagnetism,2001PhLA..292...49M,2002CQGra..19...39I,2006AnP...518..868L}. A non-standard form of gravitomagnetism has been recently analyzed by \citet{acedo2014flyby,acedo2014constraints},  in a purely phenomenological context. Eventually, the possibility of testing GM effects in a terrestrial laboratory has been considered by many authors in the past\citep{braginsky1977laboratory,braginsky1984foucault,cerdonio1988dragging,ljubivcic1992proposed,camacho2001quantum,iorio2003possibility,pascual2003telepensouth,stedman2003detectability,2006GeoJI.167..567I}; a recent proposal pertains to the use of an array of ring lasers\citep{2011PhRvD..84l2002B,ruggiero2015sagnac}, and is now underway\citep{2014CRPhy..15..866D}. 

In a recent paper \citep{2015IJMPD..2450060R},  we have investigated the gravitational field of massive rings: exploiting the GEM analogy, we have studied both the GM and the GE components of the field, produced by a thin rotating ring, orbiting the central body along a Keplerian orbit. The ring field can be dealt with as a perturbation of the background field determined by the central body. We have used a power series expansion to calculate the field in the intermediate zone between the central body and the ring. Massive rings are ubiquitous and important in astrophysics, as suggested   in \citet{2012EM&P..108..189I};  in \citet{ramos2011motion}  the effects of geometrical deformations on ring-like structures are studied, together with the implications for stability and regularity of the motion of test particles (also for  Saturn's and Jupiter's rings). Hence, motivated by the relevance of ring-like structures,  in \cite{2015IJMPD..2450060R} we have focused on the GM component of the field (the GE one is exhaustively studied in \cite{2012EM&P..108..189I}), and studied  its impact  on some gravitational effects, such as gyroscopes precession, Keplerian motion and time delay in some simplified geometric configurations.  The underlying idea is {to consider} the possibility of using these tests to estimate  the mass and  the angular momentum of  matter rings. Here, we want to pursue the study of the GM field of rotating rings: to be specific, we want to calculate the GM field in the whole space,  both in the intermediate zone between the ring and the central body and far away from the ring and central body. As for the effects of the ring field, we will focus on the perturbations of the Keplerian orbital elements of a test particle: while in the previous paper  \citet{2015IJMPD..2450060R} we have considered just the case of coplanarity between the ring and the test particle orbit, here we will consider an arbitrary configuration.  Then, we will compare the predicted secular variations with the recent observations of Solar System ephemeris \citep{fienga2011inpop10a,pitjeva2013relativistic,pitjev2013constraints}. 

The paper is organized as follows: we review the foundations of the GEM formalism in Section \ref{sec:GEM}, while in Section \ref{sec:GEMf} we obtain the GM field of the ring; in Section \ref{ssec:GEMp} we focus on the perturbations of the orbital elements determined by the GM field, and use the recent data of Solar System ephemeris to estimate the spin of ring-like structures. Conclusions are eventually in Section \ref{sec:cconc}.

\section{The GEM formalism}\label{sec:GEM}

If we  work in the \textit{weak-field and slow-motion approximation}, we may write the space-time metric in the form\footnote{Greek indices run to 0 to 3, while Latin indices run from 1 to 3; 
 bold face letters like ${\mathbf{x}}$ refer to space vectors.} $g_{\mu\nu}=\eta_{\mu\nu}+h_{\mu\nu}$, in terms of the Minkowski tensor $\eta_{\mu\nu}$ and the gravitational potentials  $h_{\mu\nu}$  which are supposed to be a small perturbation of the flat space-time metric: $|h_{\mu\nu}| \ll |g_{\mu\nu}|$. Hence, in linear approximation, on setting  $\bar h_{\mu\nu}=h_{\mu\nu}-\frac12 h \eta_{\mu\nu}$ with $h={\rm tr}(h_{\mu\nu})$, and imposing the transverse gauge condition $\bar h^{\mu\nu}{}_{,\nu}=0$, the Einstein equations take the form \citep{ciufolini1995gravitation,ohanian2013gravitation} 
 \beq
\dal \bar h_{\mu\nu}=-\frac{16\pi G}{c^4}T_{\mu\nu}\ . \label{eq:fieldgem1}
\eeq
It is a well known fact that, due to the analogy with electromagnetism \citep{2001rfg..conf..121M,2003gr.qc....11030M,Mashhoon:2005pj,mashhoon1993gravitational,2001LNP...562...83M,Bini:2008cy}, the solution of the field equations (\ref{eq:fieldgem1}) can be written in the form\footnote{Here and henceforth we use the convention  introduced by Mashhoon \citep{mashhoon1993gravitational} to exploit the standard results of electrodynamics to describe gravity in post-Newtonian linear approximation. Other conventions are used elsewhere \citep{ciufolini1995gravitation,ohanian2013gravitation}.}
\begin{eqnarray}
\mathrm{d} s^2&=& -c^2 \left(1-2\frac{\varphi_{}}{c^2}\right)\rmd t^2 -\frac4c ({\mathbf A_{}}\cdot \rmd {\mathbf r})\rmd t + \nonumber \\ &+& \left(1+2\frac{\varphi_{}}{c^2}\right)\delta_{ij}\rmd x^i \rmd x^j\, \label{eq:weakfieldmetric1}
\end{eqnarray}
in terms of the gravitoelectric  $\varphi_{  }$ ($\displaystyle \bar h_{00} \doteq 4\frac{\varphi_{}}{c^{2}}$ ) and gravitomagnetic  $A_{i}$ ($ \bar h_{0i}=-2 \frac{A_{ i}}{c^{2}}$) potentials, which are related to the sources of the gravitational field by
\beq
\varphi_{} (ct,\mb r)={}\int_{V}\frac{\rho_{}(ct-|{\mathbf r}-{\mathbf R}|, {\mathbf R})}{|{\mathbf r}-{\mathbf R}|}\rmd V,\label{eq:solgemphi1}
\eeq

\beq
A_{  i}(ct,\mb r)=\frac{2G}{c}\int_{V} \frac{j_{}^{i}(ct-|{\mathbf r}-{\mathbf R}|, {\mathbf R})}{|{\mathbf r}-{\mathbf R}|}\rmd V.\label{eq:solgemAi1}
\eeq
In the above equations $\rho_{}$ is the mass density and $j_{}^{i}$ is the mass current of the sources. So, we see that, besides the usual Newtonian contribution $\varphi_{}$, related to the mass of sources, there is a contribution related to the mass current of the sources. The gravitoelectric $\mb E_{}$ and gravitomagnetic $\mb B_{}$ fields  are then defined as
\beq
\mb E_{}= -\frac{1}{2c} \ppar{\mb A_{}}{t}-\mathbf \nabla \varphi, \quad  \mb B_{}= \mathbf \nabla \wedge \mb A_{} \label{eq:solgemEB1}
\eeq
For stationary sources, the equation of motion (i.e. the spatial components of the geodesics) of a test mass $m_{\mathrm{test}}$  moving with speed $\mb v$  in GEM fields $\mb E_{}, \mb B_{}$ turns out to be (see e.g. \citet{Bini:2008cy})
\beq
m_{\mathrm{test}}\frac{\rmd {\mathbf v}}{\rmd t}=-m_{\mathrm{test}}{\mathbf E_{}}-2m_{\mathrm{test}} \frac{{\mathbf v}}{c}\times {\mathbf B_{}}, \label{eq:gemforce1}
\eeq
to lowest order in $v/c$. In the convention used, a test particle of inertial mass $m_{\mathrm{test}}$ has gravito-electric charge $q_E=-m_{\mathrm{test}}$ and gravito-magnetic charge $q_B=-2m_{\mathrm{test}}$; the GEM Lorentz acceleration acting on a test particle is
\beq
\mathbf{\mathcal{A}}=-{\mathbf E_{}}-2 \frac{{\mathbf v}}{c}\times {\mathbf B_{}} \label{eq:florentz}
\eeq

\section{The gravitomagnetic field of rotating rings}\label{sec:GEMf}

\begin{figure}[here]
\begin{center}
\includegraphics[scale=.40]{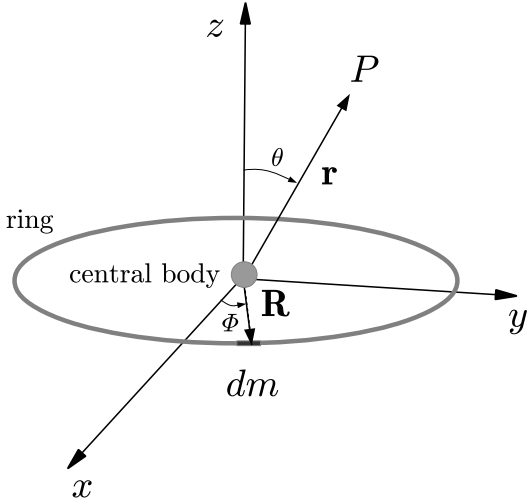}

\caption{The point $P$ has spherical coordinates $(r,\theta,\phi)$, and the origin of the coordinate system coincides with the central body; they are related to the Cartesian ones by $r=\sqrt{x^{2}+y^{2}+z^{2}}$,  $\phi=\arctan (y/x)$, $\theta=\arctan(\sqrt{x^{2}+y^{2}}/z$). The ring is in the $xy$ plane, where $\theta=\pi/2$. } \label{fig:figuraxy}
\end{center}
\end{figure}  

In this Section, we calculate the gravitomagnetic field produced by a rotating  ring; we suppose  that the ring is thin and made  of continuously distributed matter  with constant density, orbiting a central body. Furthermore, for the sake of simplicity, we assume that the ring is circular: actually, the case of an elliptically shaped ring has been considered in \citet{2015IJMPD..2450060R}, but the resulting expressions are in general unmanageable, even to lowest order in the eccentricity.

The central body is supposed to produce  its gravitational field, which is determined by its mass $M$, and angular momentum $\mb S$. In the  inertial frame where the central body is at rest, we set a  Cartesian coordinate system $\{x,y,z\}$, with the corresponding unit vectors $\mb u_{x}, \mb u_{y}, \mb u_{z}$;  if the body is located at the origin and  its angular momentum is directed along the $z$ axis, $\mb S= S \mb u_{z}$, the space-time metric to lower order has  the form (\ref{eq:weakfieldmetric1}), with $\displaystyle \varphi= \frac{GM}{r}$,  $\displaystyle \mb A =  \frac{G}{c} \frac{\left(\mb S \wedge \mb r \right)}{r^{3}}$, where $r=|\mb r|=\sqrt{x^{2}+y^{2}+z^{2}}$. In particular, the GM field turns out to be 
\beq
\mb B= \frac{G}{c} \left[\frac{3(\mb S \cdot \mb r) \mb r}{r^{5}}-\frac{\mb S}{r^{3}} \right] \label{eq:Bdipole}
\eeq
and has a typical dipole-like behaviour: in other words, it is analogous to the magnetic field produced by a dipole.

In order to evaluate the gravitomagnetic field of the rotating ring, we  proceed as follows: an infinitesimal mass element $\rmd m$ of the ring is orbiting around the central body; we know the total mass $m$ of the ring, and its angular momentum $\mb s$,  {which we assume to be \textit{constant}: in other words, we consider a stationary ring.} In our perturbative approach, we do suppose that $m\ll M$, $s \ll S$. Due to the presence of this ring, the gravitomagnetic  potential (\ref{eq:solgemAi1}) is perturbed, so that  $\mb A \rightarrow \mb A+\mb A^{\mathrm{ring}}$, where $|\mb A^{\mathrm{ring}}| \ll |\mb A|$. In particular,  we are interested in calculating this perturbation, by means of a power series expansion, (i) in the intermediate region between the central body and the ring, (ii) in the outer region of the system, i.e. away from the ring and the central body.

To this end, we consider the following geometric configuration: we suppose that the ring is in the $xy$ plane, that is the symmetry plane of the central body. In order to deal with the symmetries of the problem in a simpler way, we will also use  spherical coordinates $\{r,\theta,\phi\}$, togheter with  the corresponding unit vectors $\mb u_{r}, \mb u_{\theta}, \mb u_{\phi}$.

Let $P$ denote the point where we want to evaluate the GM field (see Figure \ref{fig:figuraxy}): its spherical coordinates are $\{r,\theta,\phi\}$ and its position vector is $\mb r$; the position vector of a mass element $\rmd m$ of the ring is $\mb R$, where $R=|\mb R|$ is the radius of the ring, and its spherical coordinates are $\{R,\pi/2,\Phi\}$.The ring uniform density is $\displaystyle \lambda= \frac{m}{2\pi R}$; furthermore, $v$ is the (constant) modulus of the mass elements speed. We must substitute $ j^{i}\rmd V \rightarrow \lambda v^{i} \rmd L$ in (\ref{eq:solgemAi1}):  $v^{i}$ are the components of the velocity, which may be written as $\mb v=-v \sin \Phi\, \mb u_{x}+v \cos \Phi\, \mb u_{y}$, and $\rmd L$ is the infinitesimal arc length of the ring. Accordingly, we get

\beq
\mb A^{\mathrm{ring}} =\frac{2G}{c}\int_{L} \frac{\lambda \left( -v \sin \Phi\, \mb u_{x}+v \cos \Phi\, \mb u_{y} \right) \rmd L}{|{\mathbf r}-{\mathbf R}|} \label{eq:deltaA0}
\eeq
We may write $\rmd L = R\, \rmd \Phi$; hence, on introducing the angular momentum per unit mass $\sigma=v\,R$, the above integral can be written as
\beq
\mb A^{\mathrm{ring}} =\frac{2G\lambda \sigma}{c}\int_{\Phi=0}^{\Phi=2\pi} \frac{ \left( - \sin \Phi\, \mb u_{x}+ \cos \Phi\, \mb u_{y} \right) \rmd \Phi}{|{\mathbf r}-{\mathbf R}|} \label{eq:deltaA01}
\eeq
This expression can be expanded in power series: for (i) $r<R$, we expand in powers of $\displaystyle \epsilon=\frac r R$, while for(ii) $r>R$, we expand in powers of $\displaystyle \epsilon=\frac R r$. Consequently,  we may write $|{\mathbf r}-{\mathbf R}|$ in the form
\begin{eqnarray*}
(i)\,  r<R: & & {|\mathbf r}-{\mathbf R}| = R \sqrt{1-2 \frac r R \sin\theta \cos (\phi-\Phi)+\left(\frac r R \right)^{2}} \\
(ii)\, r>R: & & {|\mathbf r}-{\mathbf R}| = r \sqrt{1-2 \frac R r \sin\theta \cos (\phi-\Phi)+\left(\frac R r \right)^{2}}
\end{eqnarray*}
or
\beq
{|\mathbf r}-{\mathbf R}| = d \sqrt{1-2 \epsilon \sin\theta \cos (\phi-\Phi)+\epsilon^{2}}
\eeq
\[
\mathrm{where:} \left.\begin{array}{c}(i) r<R: d=R, \, \epsilon= r/R \\ (ii) r>R: d=r, \, \epsilon= R/r\end{array}\right.
\]

Hence, we may write
\beq
\mb A^{\mathrm{ring}} =\frac{2G\lambda \sigma}{cd}\int_{\Phi=0}^{\Phi=2\pi} \frac{ \left( - \sin \Phi\, \mb u_{x}+ \cos \Phi\, \mb u_{y} \right) \rmd \Phi}{ \sqrt{1-2 \epsilon \sin\theta \cos (\phi-\Phi)+\epsilon^{2}}} \label{eq:deltaA02}
\eeq
Because of the cylindrical symmetry, we may choose the observation point at $\phi=0$: as a consequence the $x$ component of $ \mb A^{\mathrm{ring}}$ is null, and the $y$ component is equal to $ A^{\mathrm{ring}}_{\phi}$:
\beq
A^{\mathrm{ring}}_{\phi} =\frac{2G\lambda \sigma}{cd}\int_{\Phi=0}^{\Phi=2\pi} \frac{ \cos \Phi\,   \rmd \Phi}{ \sqrt{1-2 \epsilon \sin\theta \cos (\Phi)+\epsilon^{2}}} \label{eq:deltaA03}
\eeq

As shown in \citet{jackson1999classical},  the above integral (\ref{eq:deltaA03}) can be evaluated in terms of  elliptic integrals:
\begin{eqnarray}
&& \int_{\Phi=0}^{\Phi=2\pi} \frac{ \cos \Phi\,   \rmd \Phi}{ \sqrt{1-2 \epsilon \sin\theta \cos (\Phi)+\epsilon^{2}}}=  \nonumber \\ 
&=& \frac{4}{\sqrt{1+\epsilon^{2}+2\epsilon \sin \theta}} \left[\frac{\left(2-p^{2}\right) K(p)-2E(p)}{p^{2}}\right] \label{eq:jack1}
\end{eqnarray}
where $\displaystyle p^{2}=\frac{4 \epsilon \sin \theta}{\epsilon^{2}+1+2\epsilon \sin \theta}$ and $K(p)$, $E(p)$ are the complete elliptic integrals of first and second kind. For $\epsilon \ll 1$ (that is for $R \ll r$ or $r \ll R$) the result of the integral in Eq. (\ref{eq:jack1}) becomes $\displaystyle \frac{\pi \epsilon \sin \theta }{\left( 1+\epsilon^{2}+2\epsilon \sin \theta\right)^{3/2}}$ and, consequently, we may write the gravitomagnetic potential in the form
\beq
A^{\mathrm{ring}}_{\phi} = \frac{2G\lambda \sigma}{cd} \frac{\pi \epsilon \sin \theta }{\left( 1+\epsilon^{2}+2\epsilon \sin \theta\right)^{3/2}}\label{eq:jack2}
\eeq

If we perform a power-series expansion we obtain:
\begin{eqnarray*}
(i)\,  r\ll R:    A^{\mathrm{ring}}_{\phi} &=& {\frac {Gs\sin  \theta  }{{cR}^{2}}}\frac{r}{R}-3\,{\frac {Gs
  \sin^{2}  \theta    }{c{R}^{2}}}\frac{r^{2}}{R^{2}}+ \\ \nonumber &+&  \left[\frac{15}{2}\,{\frac {Gs  \sin^{3}  \theta {}^{}
}{c{R}^{2}}}-\frac 3 2 \,{\frac {Gs\sin  \theta  }{c{R}^{2
}}}\right]\frac{r^{3}}{R^{3}}+ \nonumber \\ &+& O\left(\frac{r^{4}}{R^{4}} \right) \label{eq:Aint1}
\\
(ii)\, r \gg R:  A^{\mathrm{ring}}_{\phi}&=& {\frac {Gs\sin  \theta  }{c{r}^{2}}}-3\,{\frac {Gs
  \sin^{2}  \theta    }{c{r}^{2}}}\frac{R}{r} + \nonumber \\ &+&  \left[\frac{15}{2}{
\frac {Gs{}^{}\sin^{3}  \theta }{c{r}
^{2}}}-\frac 3 2 \,{\frac {Gs{}^{}\sin\theta }{c{r}^{2}}}\right]\frac{R^{2}}{r^{2}}+ \nonumber \\ &+&  O\left(\frac{R^{4}}{r^{4}} \right) \label{eq:Aext1}
\end{eqnarray*}

According to Eq. (\ref{eq:solgemEB1}), the corresponding gravitomagnetic field can be obtained from $\mb B^{\mathrm{ring}}= \mathbf \nabla \wedge  \mb A^{\mathrm{ring}}$.  For $  r\ll R$, the gravitomagnetic field has the following components
\begin{eqnarray}
B^{\mathrm{ring}}_{r}  &=&  \frac{2Gs}{cR^{3}}\cos \theta-\frac{9Gs}{cR^{3}}\cos \theta \sin \theta \frac r R + \nonumber \\ &+& \frac{Gs}{cR^{3}} \cos \theta \left(30 \sin^{2} \theta-3 \right)\frac{r^{2}}{R^{2}}+O\left(\frac{r^{3}}{R^{3}}\right) \label{eq:Bir}
\end{eqnarray}
\begin{eqnarray}
B^{\mathrm{ring}}_{\theta}  &=&  -\frac{2Gs}{cR^{3}}\sin \theta+\frac{9Gs}{cR^{3}} \sin^{2} \theta \frac r a + \nonumber \\ & +& \frac{Gs}{cR^{3}} \sin \theta \left(6 -30 \sin^{2} \theta \right)\frac{r^{2}}{R^{2}}+O\left(\frac{r^{3}}{R^{3}}\right) \label{eq:Bitheta}
\end{eqnarray}
While, for $  r\gg R$, we obtain 
\begin{eqnarray}
B^{\mathrm{ring}}_{r}  &=&  \frac{2Gs}{cr^{3}}\cos \theta-\frac{9Gs}{cr^{3}}\cos \theta \sin \theta \frac R r + \nonumber \\ &+& \frac{Gs}{cr^{3}} \cos \theta \left(30 \sin^{2} \theta-3 \right)\frac{R^{2}}{r^{2}}+O\left(\frac{R^{3}}{r^{3}}\right) \label{eq:Ber}
\end{eqnarray}
\begin{eqnarray}
B^{\mathrm{ring}}_{\theta}  &=&  \frac{Gs}{cr^{3}}\sin \theta-\frac{6Gs}{cr^{3}} \sin^{2} \theta \frac R r + \nonumber \\ &+& \frac{Gs}{2cr^{3}} \sin \theta \left(45 \sin^{2} \theta-9 \right)\frac{R^{2}}{r^{2}}+O\left(\frac{R^{3}}{r^{3}}\right)  \label{eq:Betheta}
\end{eqnarray}


We notice that, to lowest approximation order, the gravitomagnetic field has the following expressions
\begin{eqnarray}
(i)\,  r\ll R:   \mb B^{\mathrm{ring}}&=& \frac{2Gs}{cR^{3}}\left(\cos \theta \mb u_{r}- \sin \theta \mb u_{\theta}\right)= \nonumber \\  &=& \frac{2G}{cR^{3}}\mb s \label{eq:Bint1}
\\
(ii)\, r \gg R:    \mb B^{\mathrm{ring}} &=&\frac {Gs}{cr^{3}} \left( 2 \cos \theta  \mb u_{r}+\sin \theta  \mb u_{\theta} \right)= \nonumber \\ &=& \frac{G}{c} \left[\frac{3(\mb s \cdot \mb r) \mb r}{r^{5}}-\frac{\mb s}{r^{3}} \right] \label{eq:Bext1}
\end{eqnarray}
The expression (\ref{eq:Bint1}) of the field inside the ring is in agreement with the one obtained in \citet{2015IJMPD..2450060R}; on the other hand, we see that the expression (\ref{eq:Bext1}) of the field outside the ring is, as expected,  the usual dipole field. In the following Section we are going to use these expressions to calculate the perturbing acceleration on the motion of orbiting bodies and, then, the corresponding variations of the orbital elements.

\section{Gravitomagnetic perturbations} \label{ssec:GEMp}

In this  Section we evaluate the impact of the GM of the rings on the orbital elements of test particles: in particular, we  consider below the effects on Solar System bodies. To this end,  we use the expressions of the GM field (\ref{eq:Bint1}) and (\ref{eq:Bext1}) to calculate the perturbing acceleration
\beq
\bds{W}=-2 \frac{{\mathbf v}}{c}\times { \mathbf B^{\mathrm{ring}}} \label{eq:aring}
\eeq
 then, we can evaluate its effects on planetary motions using the Gauss equations for the variations of the elements, which enable us to study the perturbations of the Keplerian orbital elements due to a generic perturbing acceleration.

\begin{figure}[here]
\begin{center}
\includegraphics[scale=.40]{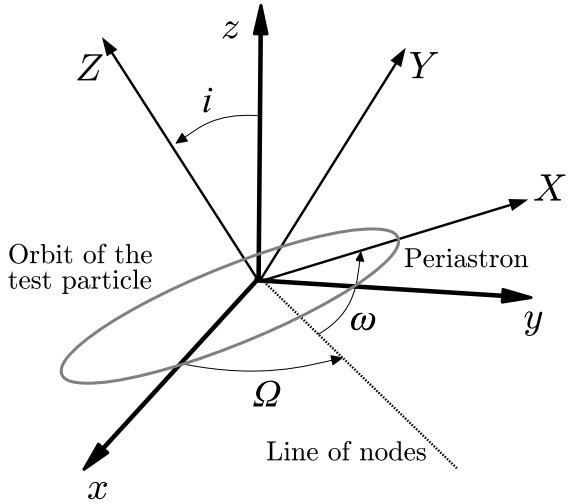}

\caption{Unperturbed orbit of the test particle} \label{fig:orbita}
\end{center}
\end{figure}

{To begin with, let us describe the configuration of the unperturbed test particle orbit. We refer to Figure \ref{fig:orbita}: besides the already mentioned Cartesian coordinate system $\{x,y,z\}$,  we introduce another Cartesian coordinate system $\{X,Y,Z\}$, with the same origin. The ring lies in the $xy$ plane, while the the orbital plane of the test particle  $XY$ plane. In particular,  we denote with $\Omega$  the angle between the $x$ axis and the line of the nodes, while the angle between the $z$ and $Z$ axes is $i$. The periastron is along the $X$ axis, and we denote by $\omega$ the argument of the periastron, i.e. the angle between the line of nodes and the $X$ axis.\\
We use the standard approach to the perturbation of orbital elements (see e.g. \citet{2003ASSL..293.....B}, \citet{2005ormo.book.....R}): to this end, we  calculate the the
radial, transverse (in-plane components) and  normal (out-of-plane
component) projections of the perturbing acceleration
 (\ref{eq:aring}) on the orthonormal frame comoving with the particle, and then we use the Gauss equations for the
variations of the semi-major axis $a$, the eccentricity $e$, the inclination $i$, the longitude of the ascending node $\Omega$, the
argument of pericentre $\omega$ and the mean anomaly $\mathcal{M}$ (see e.g. \citet{2005ormo.book.....R}). We want to stress that, as we said before, the ring is assumed to be \textit{stationary}:  this amounts to saying that  the motion of the ring matter is constant during the particle's timescale.\\
We obtain the following results.}

For the test particles orbiting outside the ring, on using the expression (\ref{eq:Bext1}) of the GM field, we have non null secular variations only for the argument of periastron and the node:
\beq
<\dot{{\omega}}>= -\frac{6Gs}{c^{2}} \frac{1}{a^{3}\left(1-e^{2} \right)^{3/2}} \label{eq:Deltaomegaext}
\eeq

\beq
<\dot{{\Omega}}>= \frac{2Gs}{c^{2}} \frac{1}{a^{3}\left(1-e^{2} \right)^{3/2}} \label{eq:DeltaOmegaext}
\eeq
Eventually, for the  longitude of the pericenter we have
\beq
<\dot{{\varpi}}>= -\frac{2Gs\left(3\cos i-1 \right)}{c^{2}} \frac{1}{a^{3}\left(1-e^{2} \right)^{3/2}} \label{eq:Deltaomegabarext}
\eeq
On the other hand, for test particles orbiting inside the ring, on using the expression (\ref{eq:Bint1}), we have the following non null secular variations:
\beq
<\dot{{\Omega}}>= \frac{2Gs}{c^{2}R^{3}} \label{eq:DeltaOmegaint}
\eeq

\beq
<\dot{{\mathcal{M}}}>=n+\frac{2Gs \cos i}{c^{2}R^{3}}\left(4\sqrt{1-e^{2}}+1 \right)\label{eq:DeltaMint}
\eeq
Moreover, for the  longitude of the pericenter we have
\beq
<\dot{{\varpi}}>=\frac{2Gs}{c^{2}R^{3}} \label{eq:Deltaomegabarint}
\eeq
{We remember that fot an
unperturbed Keplerian ellipse in the gravitational field of a body with mass $M$,  it is ${n}=\sqrt{GM/a^3}$.}

The above results can be used to make a comparison with the recent observations  \citep{fienga2011inpop10a,pitjeva2013relativistic,pitjev2013constraints}: for instance, on using the  available supplementary advances   $\Delta \dot{{\varpi}}$, we may   give estimates on the spin of ring-like structures in the Solar System.  Let us start from planets orbiting \textit{outside} the ring; in particular, we consider a hypothetical ring of matter, inside the orbits of Mars or Mercury. We obtain the following expressions for $<\dot{{\varpi}}>$, on taking into account the orbit of Mars ($e=9.34 \times 10^{-2}$, $i=1.84$, $a=1.52$ AU, \citep{ephemerides}):
\beq
<\dot\varpi>_{\mathrm{Mars}}=-7.69 s \times 10^{-28}  \mathrm{mas} \ \mathrm{cty^{-1}} \label{eq:omegabarmars}
\eeq
while for Mercury ($e=2.05 \times 10^{-1}$, $i=7.00$, $a=3.87 \times 10^{-1}$ AU, \citep{ephemerides}):
\beq
<\dot\varpi>_{\mathrm{Mercury}}=3.44 s \times 10^{-26}  \mathrm{mas} \ \mathrm{cty^{-1}} \label{eq:omegabarmercury}
\eeq
On using the data obtained by \citet{fienga2011inpop10a}, $\Delta \dot{{\varpi}}_{\mathrm{Mars}}=-0.04 \pm 0.15$ mas cty$^{-1}$; we obtain  $s \leq 1.43 \times 10^{26}$ kg m$^{2}$ s$^{-1}$. As for Mercury, it is \citep{fienga2011inpop10a}  $\Delta \dot{{\varpi}}_{\mathrm{Mercury}}=0.4 \pm 0.6$;  similarly, we obtain $s \leq 2.90 \times 10^{25}$ kg m$^{2}$ s$^{-1}$. 

As for planets orbiting \textit{inside} the ring,  we see that $<\dot\varpi>$ in Eq. (\ref{eq:Deltaomegabarint}) is constant,  so that it does not depend on the orbit of the test particle (even though the orbit is not in the plane of the ring): as for its magnitude, we obtain
\beq
<\dot\varpi>=s \, 2.9 \times 10^{-43} \left(\frac{1 \mathrm{AU}}{R} \right)^{3} \, \mathrm{mas} \ \mathrm{cty^{-1}}
\eeq
As a consequence,  the spin magntiude $s$ of a hypothetical ring at $R=1$ AU can be constrained by using the data of Venus, measured by \citet{fienga2011inpop10a}:  $\Delta \dot{{\varpi}}_{\mathrm{Venus}}=0.2 \pm 1.5$ mas cty$^{-1}$;  we obtain $s \leq 5.9 \times 10^{42}$ kg m$^{2}$ s$^{-1}$. Similarly, we can constrain the  spin of the minor asteroids belt between Mars and Jupiter, by considering $R=2.8$ AU, and using the perihelion of Mars, measured by \citet{fienga2011inpop10a} $\Delta \dot{{\varpi}}_{\mathrm{Mars}}=-0.04 \pm 0.15$ mas cty$^{-1}$;  we obtain  $s \leq 8.3 \times 10^{42}$ kg m$^{2}$ s$^{-1}$. \\

It is important to explain the meaning of the above estimates: indeed, they should be considered just as upper limits, useful to evaluate the order of  magnitude of the  effects. In fact, in actual physical situations, the GM perturbations due to the ring are present together with other effects, such as the Lense-Thirring, the $J_{2}$ effects of the central body and the Newtonian or GE effects of the ring; {among the latter, there are the tidal interactions: in particular, it is possible to show that for actual physical situations, the impact of the tidal interactions are much greater than the GM perturbations due to ring.{\footnote{A rough estimate of the ratio of the magnitudes of the GM acceleration $\mb W_{GM}$ (due to a ring of mass $m_{r}$ orbiting at distance $R_{r}$ from a central body of mass $M$) to the tidal acceleration $\mb W_{T}$ (due to a planet of mass $m_{p}$ orbiting the same central body at distance $R_{p}$) on test particle orbiting at distance $r$ ($r<R_{r}, r<R_{p}$, can be written in the form $\displaystyle \frac{W_{GM}}{W_{T}} \simeq \frac{\frac{GM}{c^{2}r}\frac{m_{r}}{R_{r}^{3}}}{\sqrt{\frac{r}{R_{r}}}\frac{m_{p}}{R^{3}_{p}}}$, which suggests that the post-Newtonian GM effects are very small.}}}
Moreover, in the case of planets orbiting outside the ring we notice that the expression of the GM field (\ref{eq:Bext1}) is the same as the GM field of the central body, in terms of its own angular momentum (\ref{eq:Bdipole}); in particular, the secular variations are the same as those of the classical Lense-Thirring effect \citep{2001NCimB.116..777I}. As a consequence, far away from the central body and the ring, the total GM field will depend on the sum of the angular momenta of the central body \textit{and} the ring,  and it would be very difficult (at least for the chosen ring configuration) to set constraints on the angular momentum of the ring. 



\section{Conclusions}\label{sec:cconc}

In this paper we have focused on the GM field produced by rotating rings of matter, orbiting  around a central body, regarded as a small perturbation of the leading gravitational field of the central body. In particular, we have considered a thin circular ring, with constant matter density, and calculated its field, in the form of power law,  in the intermediate zone between the central body and the ring, and also far away from the ring and the central body.  Then, we have  used the lowest order expression of the GM field, both inside and outside the ring, to calculate  the corresponding perturbing acceleration on the Keplerian orbit of a test particle, with arbitrary inclination with respect to the ring plane, thus extending some previous results. As a possible application, we have evaluated the impact of the GM perturbations on  the Keplerian orbital elements, to make a comparison with the available data in the Solar System: namely, on taking into account the data of the planetary ephemeris, we have used the predicted perturbations of the orbital elements to give rough estimates on the spin of  ring-like structures. These results are preliminary: the simple model that we have considered, in fact, can be used to obtain upper limits on the spin of the rings, since we have not taken into account the other perturbations that are present. However we suggest that, at least in principle, by means of a more realistic and systematic analysis of the perturbations, it could be possibile to infer more information on the spin of  ring-like structures. 


\bibliographystyle{spr-mp-nameyear-cnd}  
\bibliography{bibo}                

\end{document}